# Water and the Many-Body Imagination

Albert Libchaber[1], Tsvi Tlusty[2]

[1]Rockefeller University, New York, USA; [2]UNIST, Ulsan, Korea

An electron crosses the cold Fermi sea. The sea recoils, screens, remembers. It returns the electron dressed: a quasiparticle with a mass, a lifetime, and a Green function. This was Nozières' lesson. We carry it to water. Electrons become dipoles; the Fermi sea becomes a hydrogen-bonded polar liquid. We ask the same question: can the collective modes of water live long enough and reach far enough for molecular machines to interact and synchronize?

## A bouquet of ideas

At the École Normale Supérieure there was a special group, headed by Pierre Aigrain, studying semiconductor physics. It was a very special, unconventional group, with Julien Bok and Philippe Nozières as the two theoretical physicists. Philippe was already world famous, having published in 1963 a book on many-body theory (*Le problème à N corps: propriétés générales des gaz de fermions* [1]), the first of its kind, with the already famous Green function, called by the users the green plague. It was a gate into the many-body problem.

I was the experimental physicist, educated by Philippe in many-body theory. Later, I carried this education elsewhere, developing a lot of nonlinear experiments, up to the period-doubling cascade and the onset of chaos.

Julien Bok was working on semiconductor applications and new devices; Philippe, of course, on many-body theory. This huge diversity of subjects was the hallmark of Pierre Aigrain, in contrast to Alfred Kastler's focus on atomic physics or Yves Rocard's on natural phenomena, and Anatol Abragam's on nuclear physics at Saclay. The other famous theory group was led by Pierre Gilles de Gennes at the École de Physique et Chimie, where physics entered chemistry in a very original manner.

Kastler, Abragam, and Rocard belonged to the generation that rebuilt French physics. De Gennes, Cohen-Tannoudji, and Nozières were the new generation, after the terrible period under Prince

Louis de Broglie, during which physics was somewhat incoherent. At the École Normale we had Aigrain, who had new ideas every day, and Philippe, who tested their limits. Aigrain opened the possible; Philippe found the limit. What an atmosphere. What a creative spirit. There were constant exchanges with the Russian Landau school, and the feeling that originality was still possible. Much later, by another path, the same freedom would reappear in Yves Couder and his great classical quantum experiment, where he observed walking and orbiting droplets [2].

This is Albert Libchaber's reminiscence.

It is in this spirit that we turn now to water. We believe that a new theory of water belongs naturally to Philippe's many-body world. Long treated as the background of biophysics, water is becoming one of its central elements.

**Water after Nozières**

Nozières asks what is collective: which modes persist, propagate, remember, and bind distant events. The elementary object is the object together with the world it disturbs; its Green's function is the medium's memory of a local action. With this lesson, we come to water.

Water hides its strangeness in plain sight. We drink it, swim in it, and forget how singular it is. To think about water, one needs to descend to the atomic scale and climb back to the collective. What then appears at each scale is the coexistence of seemingly incompatible traits. We follow this route in three settings: first in bulk water, then near surfaces and in reduced dimension, where geometry opens new modes; and finally, we ask how this affects molecular machines, where these modes may enter function itself.

Water begins with oxygen, a small, strongly electronegative atom that binds two hydrogens [3]. Together, they form a small V with a dipole of ~0.06 e·nm – not exceptionally strong yet a very compact dipole. A cubic nanometer contains thirty-three such V's. Were they to point together they would carry a polarization density of ~2 $e/nm^2$, on the scale of a ferroelectric solid. Thermal motion randomizes the orientation of dipoles, but fields and surfaces can recruit it, as they do around molecular machines. A field gives it bias; a surface gives it geometry; a molecular machine gives it timing.

From the molecular scale follows the mesoscopic. The compact $H_2O$ molecules pack many hydrogen bonds into little volume and area, making water highly cohesive. A water surface costs $\gamma \sim 20\ k_BT/nm^2$ in disrupted hydrogen bonds. Yet, the compactness keeps the liquid mobile: small molecules, fast reorientation, low viscosity, $\eta \sim 1$ cP. The ratio is a large velocity, $\gamma/\eta \sim 80$ m/s. A 1 nm wrinkle on a water surface smooths out in ~10 ps. Water makes a costly surface but moves it quickly.

The agility and compactness of water show in its electric response. In vacuum, two elementary charges 1 nm apart cost ~60 $k_BT$. In water, this cost falls to ~0.7 $k_BT$: Coulomb energy brought down to thermal scale. This reduction is because many dipoles are crowded into little volume and turn quickly; hydrogen bonds switch, neighbors exchange. In about 8 ps, bulk polarization relaxes, giving water its low-frequency dielectric constant, $\varepsilon \sim 80$.

**$H_2O$ plus geometry**

With this in mind, we turn to interfaces and reduced dimensions. In bulk, the polarization has no preferred direction; it is dispersed among fluctuations. A surface imposes a normal, interrupts the hydrogen-bond geometry, and selects the modes of response. Confinement suppresses some fluctuations and amplifies others. Boundary and distance enter the Green function. The physics is now $H_2O$ plus geometry.

From this union comes a world of singularities joining angstroms to meters. Consider a drop of water spreading on a surface. From afar, it is a smooth lens of water, ending at a sharp contact angle. But follow the wedge of water down, as Hervet and de Gennes did, and it opens into new structure [4]. Ahead of the visible contact line runs a nanometric film, a precursor laid on the solid before the drop arrives. At the level of dimensional scaling, the film height is $h \sim (A/\gamma)^{1/2}$, where $A$ is the Hamaker constant. Thus, the thickness of the film is nanometric, but its reach expands much further, as $l \sim (\gamma h t/\eta)^{1/2}$. The film is born microscopic and travels into the mesoscopic.

Long before this became theory, it had already been an industrial nuisance. The liquid then was not water, but oil. In the old Bell System, the switches were electromechanical and the contacts were electrical. After a while, the electrical contacts were fouled [5]. The explanation was that a drop of oil left behind developed a propagating precursor film that altered the contacts [6]. It took time to understand. At a surface, a liquid can precede itself. The drop has an edge; the film has gone beyond it.

**New water at the boundary**

The drop has shown what a boundary can do. It does not merely correct the bulk liquid. It completely changes the object we see. A constant measured in open water becomes, near a wall, a quantity with location and direction. At a boundary, resistance divides. Water may be hard to shear within the layer and easy to slide over the wall – viscous within, slippery without. One friction belongs to water against itself; the other, to water against the solid. Under molecular confinement even this distinction begins to blur. The water and boundary become one object [7]. Likewise, dielectric screening acquires direction, and diffusion couples to charge and flow.

This is the new water revealed by nanofluidics and confinement. It can slip almost without resistance on carbon, and slow on another surface just as flat [8]. Its polarization can be weak across a slit and strong along it [9,10]. Its ions move with hydration shells and surface charge. Its hydrogen bonds can form patterns unavailable to the bulk.

The molecule is still $H_2O$. The medium now includes wall, distance, charge, lattice, and friction. These are not external conditions. They write hydrodynamics and electrodynamics into the same law: slip coupled to surface charge, flow to ionic current, screening to confinement, polarization to friction.

**The body and its wake**

Before we turn to living machines, we return once more to the basic lesson of many-body physics: an object is changed by the medium it disturbs. Eiffel saw this first in bodies dropped from his Tower [11]. Later, measuring spheres in a wind tunnel [12], he found that as the speed increased, the drag coefficient suddenly fell. Nothing had changed in the body. The flow had changed: the boundary layer, the point of separation, the wake. A small change near the surface altered the force felt far away.

Thus the measured object in Eiffel's experiment was the body together with its boundary layer and wake. Nozières gave the microscopic form of this lesson; Eiffel, the macroscopic. Between them lies the mesoscopic body of the protein.

## Living machines in water

A protein is a few nanometers across. At this scale, water meets a dynamic boundary: charged, polar, curved, hydrophobic in patches, cut by clefts and hinges. This boundary is not fixed. During turnover, atoms move, charges shift, protonation states may change, cavities open and close. Each change requires the surrounding water to rearrange: to screen, reorient, flow. This response is a finite cloud of water governed by the machine [13]. Such machines are never isolated; they belong to a cascade of machines, nested across scales, that forms living matter [14].

For one machine to affect another through water, their clouds must overlap; the overlap must survive long enough through a functional action; and the interaction energy must be large enough to bias the action. The first demand is modest. A cubic micrometer of bacterial cytoplasm contains millions of protein molecules. The crowded proteins are separated by water gaps of only a few nanometers. At membranes and in molecular assemblies, the gaps are smaller still [15].

## The orientational field

Nanometers and microseconds are typical scales for protein machines, but too large for the correlated response of bulk water. Charge is screened within less than a nanometer, the Debye length. Polarization is lost in picoseconds, the Debye relaxation scale of bulk water. A response that reaches several nanometers and lasts toward a microsecond cannot be the ordinary bulk response. If such a response exists, its origin must be sought at the interface.

Nozières gives the dressed object. The electron is dressed by the Fermi sea, the protein by water. Abrikosov gives the language for the field that dresses it. In a superconductor, the order parameter is complex and its phase is a physical field. Phase gradients cost energy, vortices organize space, and boundaries select phase patterns [16].

Interfacial water has no quantum phase. Its physical variable is polarization. We describe the tangential response by a vector field $\mathbf{p}(\mathbf{r})$, whose 2D direction gives a phase. Charges, hydrogen bonds, and geometry anchor the direction. Thus the protein becomes a boundary condition on the field.

**The Green function of proteins in water**

Within this view, the question is the Green function $g(\mathbf{r})$ of the polarization field $\mathbf{p}(\mathbf{r})$: once a protein disturbs the water, how far does the disturbance reach, how long does it remain, and how does another boundary feel it? What follows is a tentative scaling argument.

Consider an active protein. Its action on the surrounding water is summarized by an effective surface field $\mathbf{E}(\mathbf{r})$, generated by charge, hydrogen bonds, and polarity. The interfacial water responds by polarizing along the surface, $\mathbf{p}(\mathbf{r}) \sim \varepsilon_0\chi \int g(\mathbf{r} - \mathbf{r}')\, \mathbf{E}(\mathbf{r}')\, d^2\mathbf{r}'$. The susceptibility $\chi$ sets the magnitude of the response, and the Green function g(r) sets its reach, $\xi$ [17].

The range $\xi$ is set by a balance between the cost of polarizing water and the cost of making neighboring molecules point differently. On a molecular scale, $\xi \sim a(J/M)^{1/2}$ where $a \sim 0.3$ nm is the spacing between water molecules, $M \sim 1/\chi$ is the local polarization cost, and $J$ is the cost of misaligning neighbors. In bulk water, $J$ and $M$ are both thermal in scale so $\xi$ extends over only a few molecules. A local disturbance may be strong, but soon loses its direction.

At an interface, the balance changes. In water confined to a few molecular layers, the in-plane susceptibility $\chi$ may rise from 80 toward $10^3$ [10], driven by a strongly disordered hydrogen-bond network. One may speculate that $\xi$ is lengthened, carrying a sub-nanometer bulk range toward a few nanometers at the interface. The effect is therefore modest but enough to allow interaction between crowded protein machines.

**Slipping and synchronization**

The discussion so far has been static. A protein machine, however, is dynamic and cyclic. Its field $\mathbf{E}(\mathbf{r},t)$ changes, and the polarization $\mathbf{p}(\mathbf{r},t)$ changes with it. At an interface, motion acquires a possibility absent from bulk water: it can slip along the boundary. Viscosity carries the motion within the layer; weak friction against the surface allows it to travel.

For a layer of thickness $h$ and slip length $b$, the hydrodynamic range scales as $\ell \sim (b\, h)^{1/2}$ [18]. If $h \sim 1$ nm and $b$ were ~10 nm, $\ell$ would be several nanometers, comparable to the spacing between crowded proteins. The slip length near a protein is unknown, but if interfacial friction is sufficiently weak, the disturbance may travel farther along the interface. Thus, there are two ranges, of polarization, $\xi$, and of hydrodynamics $\ell$. The first is set by susceptibility and orientational stiffness; the second, by viscosity and friction. A working protein couples them, because the motion it induces both reorients and displaces the surrounding water [19].

Time enters through the dynamic Green function, $g(\mathbf{r},\omega) \sim g(\mathbf{r})/(1 - i\omega\tau)$. The relaxation time $\tau$ is less constrained. Confinement can slow water and ionic motion, but whether the interfacial response retains memory on protein timescales is precisely what remains to be established. For $\omega\tau \lesssim 1$, the response retains much of its magnitude; near the crossover, it also acquires a phase lag. Transition rates depend exponentially on activation free energy. A shift of even a fraction of $k_B T$ can therefore change the rate, and with it the progress of a protein through its cycle [20]. If interfacial slip carries this disturbance across the gap, neighboring protein machines may become dynamically coupled – and, under suitable conditions, synchronize.

Thus, the many-body imagination with which we began now rests on one object: the Green function of water around active proteins. Theory must give it form; experiment, its reach and memory. If the two meet, we shall have touched something elementary in the physics of life.

**Acknowledgments.** T.T. acknowledges support from the National Research Foundation of Korea under Grant No. NRF-RS-2025-00573354, the InnoCORE Bio-MAX program, and the U.S. Office of Naval Research under Award No. N00014-26-1-2401.